\documentclass[extra,mreferee]{gji}
\usepackage{url} 
\usepackage{lineno}
\usepackage{soul}
\usepackage{graphicx}
\usepackage{amsmath}
\usepackage{booktabs}
\usepackage{autobreak}
\usepackage{multicol}
\usepackage{multirow}
\allowdisplaybreaks
\makeatletter
\newcommand{\atsymbol}{@}

\title[]  
  {Learned frequency-domain scattered wavefield solutions using neural operators} 

\author[]
  {Xinquan Huang$^1$\thanks{Now at: University of Pennsylvania, Philadelphia, PA 19104, USA.} and Tariq Alkhalifah$^1$ \\
  $^1$ Physical Science and Engineering division, King Abdullah University of Science and Technology, \\
  Thuwal \emph{23955-6900}, Saudi Arabia
  }

\begin{document}

\maketitle

\begin{summary}
Solving the wave equation is essential to seismic imaging and inversion. The numerical solution of the Helmholtz equation, fundamental to this process, often encounters significant computational and memory challenges. We propose an innovative frequency-domain scattered wavefield modeling method employing neural operators adaptable to diverse seismic velocities. 
The source location and frequency information are embedded within the input background wavefield, enhancing the neural operator's ability to process source configurations effectively.
In addition, we utilize a single reference frequency strategy, which enables scaling from larger-domain forward modeling to higher-frequency scenarios, thereby improving our method's accuracy and generalization capabilities for larger-domain applications. 
Several tests on the OpenFWI datasets and realistic velocity models validate the accuracy and efficacy of our method as a surrogate model, demonstrating its potential to address the computational and memory limitations of numerical methods. 
\end{summary}


\section{Introduction}
Solving wave equations is pivotal for various geophysical applications like reverse time migration and full waveform inversion. 
However, the computational and memory bottlenecks often hinder real-time applications like microseismic imaging for CO2 monitoring. 
Frequency-domain simulation \citep{Marfurt1984} offers a reduced dimension solution space for multi-scale waveform inversion. 
Yet, it still struggles with exponentially increasing memory demands with increasing model size and the necessity for repeated simulations across various frequencies.

With the development of modern machine learning techniques comes the possibility of using neural networks to surrogate numerical modeling.
Physics-informed neural networks (PINNs) have recently been employed for time-domain and frequency-domain wavefield modeling, demonstrating their potential in solving complex wave equations in irregular geometry through meshless solutions \citep{moseley_solving_2020,voytan_wave_2020,Alkhalifah2021Wavefield,song_versatile_2021,huang_pinnup_2022,zheng_ground-penetrating_2023,sethi_hard_2023,wu_helmholtz-equation_2023}. 
Despite their multi-source and multi-frequency capabilities \citep{huang_high-dimensional_2022}, PINNs' need for retraining or transfer learning for new velocities limits their generalization. 
Consequently, developing neural networks for a range of parametric PDEs becomes crucial.
Neural operators \citep{li_fourier_2020} have shown great potential in solving for parametric PDEs, which provide instant PDE solutions given various coefficients and boundary conditions. 
Recently, they have also been used for seismic wavefield simulation, e.g., 2D/3D time-domain wavefield simulation \citep{yang_rapid_2023,lehmann_fourier_2023,zhang_learning_2023,lehmann_3d_2024}. 
However, learning seismic waveform modeling in the time domain faces several challenges: (1) Learning the mapping directly from input 2D velocities and source locations to 3D wavefield volumes (spanning 2D spatial and 1D time-lapse domains) requires huge memory cost \citep{zou_deep_2023}; (2) Learning the mapping between the snapshots at different times within recurrent composition though decreases the memory cost, will be subject to error propagation \citep{huang_lordnet_2024} as well as the relatively huge inference cost due to the iterative nature. 
Thus, learning in the frequency domain using neural operators, which avoids the above limitations, could efficiently handle parametric wave equations. 

Although current neural operator implementations for the frequency domain wavefield solutions offer memory efficiency, they typically often use the end-to-end framework to learn the mapping from the velocities, source locations, and frequencies to the corresponding wavefield solutions without additional feature engineering \citep{zou_deep_2023,zhang_learning_2023,li_solving_2023}. 
Specifically, they treat the source locations as binary masks and frequencies as an additional channel for inputting neural operators with constant values.
However, without considering the physics of seismic wavefields, this approach offers suboptimal inputs for convolution-based feature extraction networks. Thus, it requires additional modifications to the network architecture, e.g., paralleled Fourier neural operator \citep{li_solving_2023} or limits the target frequencies of modeling \citep{zou_deep_2023} to make the neural operators handle various velocities and source configurations (e.g., locations and frequency) simultaneously.

To tackle this challenge, we propose learned frequency-domain scattered wavefield solutions using the Fourier neural operator. 
Unlike conventional methods that add source location and frequency as input channels, our proposed method implicitly embeds such information into an analytically evaluated background wavefield for a homogeneous background velocity. 
So, the neural operator focuses on mapping this background wavefield, with the frequency and source information embedded in it, to the scattered wavefield (or the full wavefield) for a given velocity model also as an input.
The proposed method allows the neural operators to simultaneously handle the modeling for various velocities, frequencies, and source locations.
Moreover, to improve generalization for larger-domain velocities beyond the training sample scope, we propose incorporating the single reference frequency \citep{huang_single_2022} to enhance the generalization ability.
We demonstrate the effectiveness of the proposed method on the OpenFWI dataset \citep{deng_openfwi_2022} and more realistic models, e.g., extracted from Overthrust models. 

\begin{figure}
    \centering
    \includegraphics[width=1.\columnwidth]{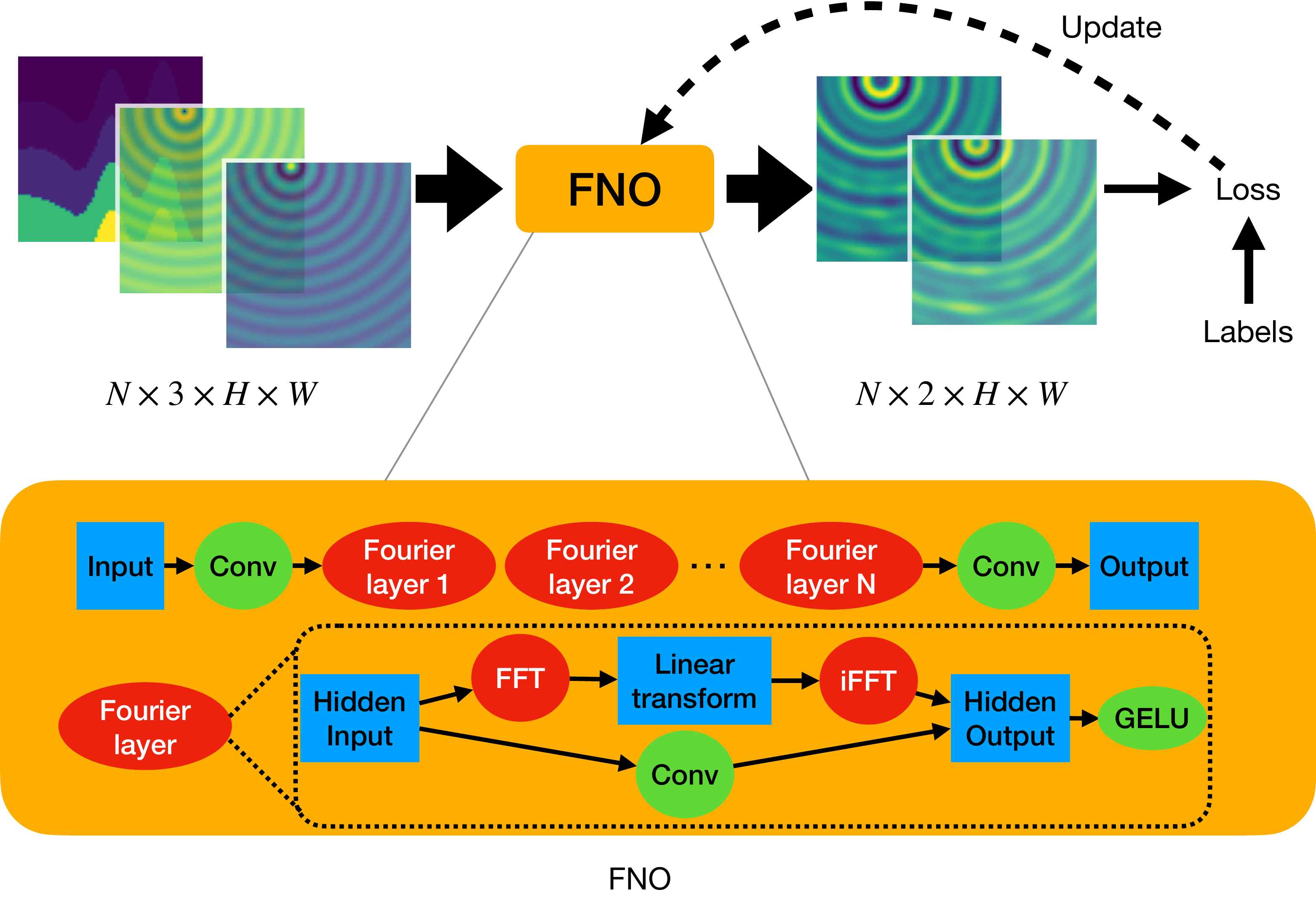}
    \caption{The pipeline of the proposed method, where the labels denote the reference results using the finite-difference method, $N$ denotes the number of samples, and $H$ and $W$ are the height and width of the images. The input has three channels, including velocity and real and imaginary parts of the analytical background wavefield, while the output consists of two channels representing real and imaginary parts of the scattered wavefield.
    For the FNO description, we refer the readers to \cite{li_fourier_2020}.}
    \label{fig:diagram}
\end{figure}

To summarize, the main contributions of our paper are as follows:
\begin{itemize}
    \item We propose to utilize analytical background wavefields to implicitly represent the source location and frequency information for neural operator-based seismic wavefield-based simulation, yielding better accuracy and generalization.
    \item We use a reference frequency strategy to enhance the performance of the trained neural operator on a larger unseen simulation domain.
    \item We validate the proposed method on a class of velocity models from OpenFWI and more realistic models and demonstrate the effectiveness of the proposed method.
\end{itemize}
\section{Methodology}
\subsection{Operator learning for Partial differential equations}
Operator learning provides a stable way to build simulations for parametric PDEs, yielding a magnitude of speed-up compared to conventional numerical simulation.
Parametric PDEs, in general, can be defined as
\begin{align}\label{eq:pdes}
    \mathcal{L}_a[u](x)&=0, \quad x\in\Omega 
\end{align}
where $a\in\mathcal{A}$ denotes the parameters of the operator $\mathcal{L}$, such as coefficient functions, and $u\in\mathcal{U}$ is the physical phenomenon we aim to solve for and belonging to solution space $\mathcal{U}$. The parameters and the solution are a function of location $x$ defined within domain $\Omega$.
For learning an operator, we assume that, for any $a\in\mathcal{A}$, there exists a unique solution $u=\mathcal{F}(a)\in\mathcal{U}$ satisfying (\ref{eq:pdes}), then $\mathcal{F}$ is the solution operator of (\ref{eq:pdes}). 
Regarding the acoustic wave equation in seismic modeling, $a$ includes the source locations, frequencies, and velocity models, and $u$ represents the corresponding simulated wavefield.

The acoustic wave equation in the frequency domain (the Helmholtz equation) is 
formulated as follows:
\begin{equation}
\left(\frac{\omega^2}{\mathbf{v}^2}+\nabla^2\right) \mathbf{U}(\mathbf{x})=\mathbf{s},
\label{equ:wave_equation}
\end{equation}
where $\omega$ is the angular frequency, $\mathbf{v}$ is the velocity, $\mathbf{U}$ is the complex frequency-domain wavefield, $\mathbf{s}$ is the source term at the location $(x_s, z_s)$.
Given a source location, a frequency, and a velocity model, we typically define a neural network $f$ with parameters $\theta$, $f_\theta(\mathbf{v}, x_s,z_s,\omega)$, to represent $\mathbf{U}$.
\subsection{Embedding the source configurations into the background wavefield}
With $f_\theta(\mathbf{v}, x_s,z_s,\omega)$ and reference pre-generated modeling results as labels, we can train the neural network $f$ to act as a surrogate modeling tool. 
However, this approach integrates the source location and the frequency as additional input channels, with a scalar value representing frequency and a binary mask denoting the source location \citep{yang_rapid_2023}. 
While structurally convenient, such representations offer limited information for effectively utilizing learned kernels in feature extraction. 
Instead, here, we propose embedding such information, including the frequency and the source location, in the background wavefield $\mathbf{U}_0$, defined, for 2D media, as:
\begin{equation}
\mathbf{U}_0(x, z)=\frac{i}{4} \boldsymbol{H}_0^{(2)}\left(\omega \sqrt{\frac{\left\{\left(x-x_s\right)^2+\left(z-z_s\right)^2\right\}}{\mathbf{v}_0^2}}\right),
\label{equ:background}
\end{equation}
where $\boldsymbol{H}_0^{(2)}$ is the zero-order Hankel function of the second kind, $(x_s,z_s)$ is the source location, $\mathbf{v}_0$ is the constant background velocity, and $i$ is the imaginary unit. 
The beauty of using background wavefield is that there is no need to specify the location of the source or the frequency while still being able to handle various source locations and frequencies.
In addition, the constant background velocity can be regarded as the zero-wavenumber component of the velocity $\mathbf{v}$. 
Thus, the background wavefield also provides guided background information for the neural network since it controls the shape of the full wavefield. 
This representation or transformation of the source location and frequencies provides rich information for convolution-based feature extraction.
Then, the input to the neural networks is changed from the $(\mathbf{v}, x_s,x_z,\omega)$ to $(\mathbf{v}, \mathbf{U}_0)$.

\subsection{Learning for the scattered wavefield}
As mentioned earlier, we can train the network to learn either the full wavefield or the residual wavefield (in our case, the scattered wavefield). 
There exists a general observation in neural operator-based fluid simulation that learning the residual of the physical wavefield rather than directly learning the physical field itself will improve the accuracy of the neural network-based simulation \citep{wandel_teaching_2021,huang_lordnet_2024}.
In addition, \cite{Alkhalifah2020} showed that solving for the scattered wavefield instead of directly learning wave equation solutions can help us avoid the point-source singularity during the training of physics-informed neural networks. 
In our case, this allows the neural operator training to focus on the relatively lower amplitude scattered wavefield. Then, the full wavefield can be obtained by simply adding the explicitly computed background wavefield. This can be thought of as a form of residual or skip connection that has demonstrated its benefits in machine learning training \citep{wandel_teaching_2021}.
Inspired by those facts and potential improvements, and considering we have a background wavefield with embedded frequency and source location information, we design the neural operator to predict the scattered wavefield, $\delta\mathbf{U}=\mathbf{U}-\mathbf{U}_0$, satisfying
\begin{equation}
\frac{\omega^2}{\mathbf{v}^2} \delta \mathbf{U}+\nabla^2 \delta \mathbf{U}+\omega^2\left(\frac{1}{\mathbf{v}^2}-\frac{1}{\mathbf{v}_0^2}\right) \mathbf{U}_0=0.
\label{equ:scattered_equation}
\end{equation}

Following the conventional supervised learning paradigm, we randomly sample source locations, frequencies, and velocities $\{v^i,x_s^i,z_s^i,\omega^i\}$ to compute the background wavefield $U_0^i$ by solving (\ref{equ:background}), and then generate the corresponding wavefield $\delta U^i$ by solving (\ref{equ:scattered_equation}) numerically, with a optimal 9-point finite difference method, for each sample. Using these samples, we train a neural network $f_\theta$ with input $(v^i,U_0^i)$ to predict the scattered wavefield (or the full wavefield by adding a residual connection from the input background wavefield directly to output) using a mean squared error loss
\begin{equation}
    L(\theta) = \frac{1}{N}\sum^N_{i=1}{\Vert f_\theta(v^i,U_0^i)-\delta U^i\Vert_2^2},
\end{equation}
where $N$ is the number of training samples.  
As for the choice of the neural network, we select the commonly used benchmark network, the Fourier neural operator (FNO) \citep{li_fourier_2020}, to test the framework. The full pipeline of the approach is shown in Figure~\ref{fig:diagram}. 
To further justify the use of the scattered wavefield as output for our implementation, we include an ablation study on various output types in the discussion (Section \ref{sec:discussion}).

\section{Numerical Examples}
We conducted tests on two velocity datasets, OpenFWI \citep{deng_openfwi_2022} and more realistic models, utilizing the Fourier Neural Operator (FNO) with four blocks (Fourier layer shown in Figure \ref{fig:diagram}). 
Each block was configured with a truncated Fourier mode set to 48 and a Fourier layer width of 128. 
The activation function used here is the Gaussian Error Linear Unit (GELU) \citep{hendrycks_gaussian_2016}.
We do not normalize the datasets like what is usually done in computer vision tasks. Instead, we directly feed the velocities in km/s and the background wavefield into the neural network. Hence, for different input pairs, they can be used without any further processing.
We employed an Adam optimizer to train the FNO, using a learning rate 1e-3. The training was conducted over 1000 epochs, with a batch size of 128 for the OpenFWI test and 64 for realistic models. The chosen frequency range, from 3 Hz to 21 Hz, aligns with the primary frequency band often used in seismic applications.

\subsection{Synthetic tests on OpenFWI dataset} 
\begin{figure}
    \centering
    \includegraphics[width=0.5\columnwidth]{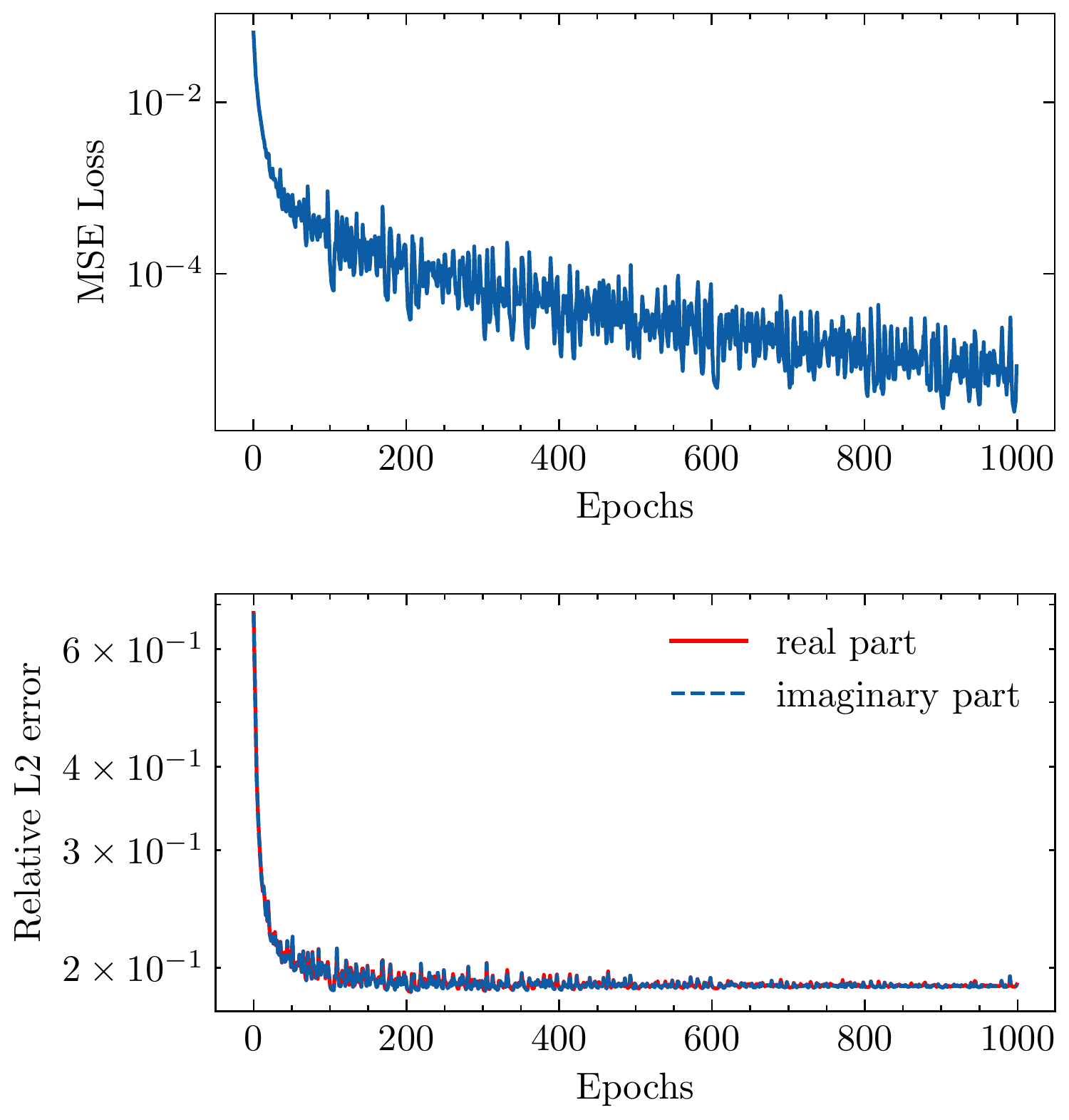}
    \caption{The training history for the 'CurveVelA' test. \textbf{Top}: the training loss curve; \textbf{Bottom}: the validation error.}
    \label{fig:training_curveA}
\end{figure}

\begin{figure}
    \centering
    \includegraphics[width=0.6\columnwidth]{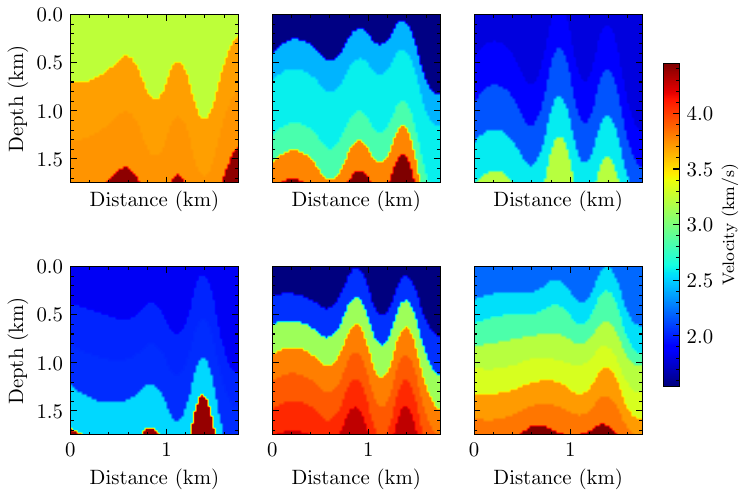}
    \caption{Samples of velocity models from the 'CurveVelA' class.}
    \label{fig:openfwi_velocities}
\end{figure}

Here, we first applied the proposed method to the OpenFWI dataset, focusing on the 'CurveVelA' velocity class \citep{deng_openfwi_2022}, as illustrated in Figure~\ref{fig:openfwi_velocities}.
This class of velocities includes curved layers, and the value of velocity within the layers gradually increases with depth.
This test involved a comprehensive training set comprising 9,000 samples, encompassing various velocities, source locations, and frequencies, and a corresponding validating set comprising 1,000 samples. 
For each sample, we randomly pick the velocity model from 'CurveVelA', randomly set the source location on the surface, randomly draw a frequency value from 3Hz to 15 Hz, and choose the background velocity of 1.5 km/s to generate the corresponding background wavefield and scattered wavefield.
The validation error is estimated using the relative L2 norm error, which equals the Euclidean distance from the prediction to the ground truth divided over the Euclidean norm of the ground truth.
Although the Fourier neural operator can learn the mapping between infinite function space, the training data here are discretized samples with a spatial resolution of 139$\times$139.
The training curve and validation error analysis are shown in Figure~\ref{fig:training_curveA}, and as we can see, the training converges well.
\begin{figure}
    \centering
    \includegraphics[width=1.0\columnwidth]{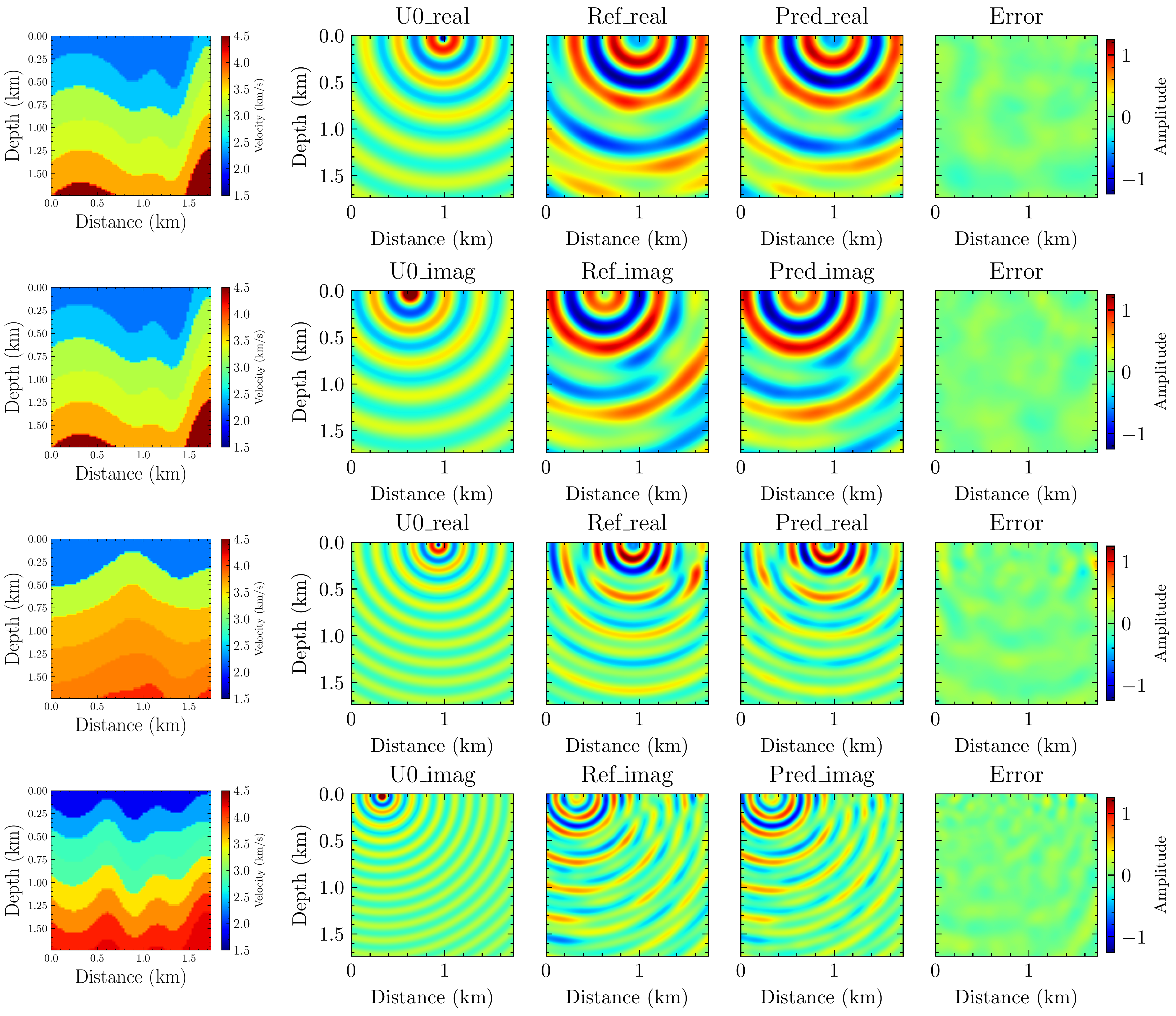}
    \caption{The results of the network predictions for unseen velocity models (Column 1) from the test set. The plots include the background wavefield (Column 2), the reference scattered wavefields from numerical solutions (Column 3), the predicted scattered wavefields (Column 4), and the difference between the predictions and the numerical solutions (Column 5). In rows 1 and 3, we show the real part of the wavefields, while in rows 2 and 4, we show the imaginary part of the wavefields. 
    The first two rows are the results of the same velocity and same frequency of 4.25 Hz but with different source locations. The third row is the result of the prediction on another velocity but with a frequency of 7.55 Hz, while the fourth row is the result of 9.88 Hz.}
    \label{fig:openfwi_results}
\end{figure}

After the training, we evaluated the FNO for various unseen velocities, shown in Figure~\ref{fig:openfwi_results}.
The frequency is randomly drawn for each case from the frequency range of 3 Hz to 15 Hz, and the source location is set on the surface and randomly drawn from the range given by our surface dimension.
For comparison, we use the optimal 9-point finite difference method to calculate the reference results with a spatial interval of 0.0125 km at both $x$ and $z$ directions.
The predictions of FNO match well with the reference scattered wavefields from numerical finite-difference modeling. 
It demonstrates that the network managed to extract the source location information and the frequency from the background wavefield and learn the relations between the various velocities and simulated complex wavefields.

\subsection{Comparison with conventional neural operator-based methods}
\begin{figure}
    \centering
    \includegraphics[width=0.9\linewidth]{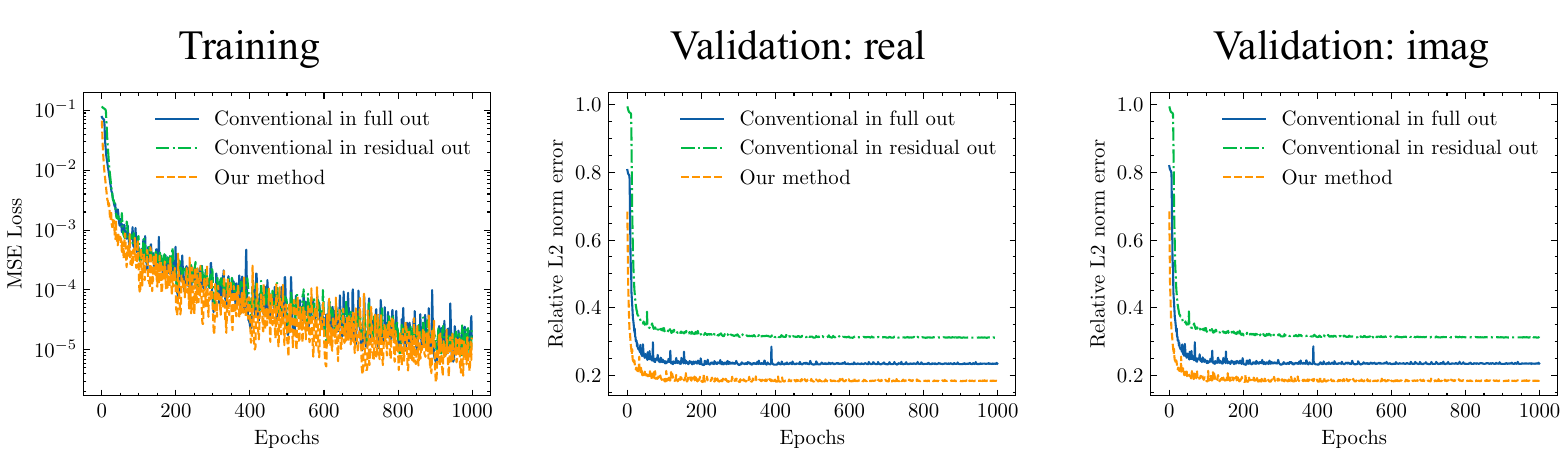}
    \caption{The comparison between different approaches on the training histories and validation errors for the real and imaginary parts.}
    \label{fig:pipeline_comp}
\end{figure}

As mentioned above, conventional neural operator-based methods treat
the source locations as binary masks and frequencies as additional channels. This approach offers suboptimal inputs
for a convolution-based feature extraction network, yielding a degraded performance. 
In this subsection, we demonstrate the weakness of conventional methods, especially when compared to our proposed method for enhancing wavefield representation.
For a fair comparison, we use the same training and validation datasets, as well as the same configurations for the FNOs, to train the neural operator using the different approaches (pipelines).
We specifically use three different approaches here: the conventional inputs with full wavefield outputs (marked as "Conventional in full out"), the conventional inputs with scattered wavefield outputs (marked as "Conventional in residual out"), and our method.
To have the evaluation metrics consistent and comparable, for the full wavefield output, the calculation of the relative L2 norm error is also done by the calculation of the L2 norm of the errors divided by the L2 norm of the scattered wavefield.
The training loss curves and the validation error curves are shown in Figure~\ref{fig:pipeline_comp}.
It is obvious that our approach, implicitly embedding the source location and frequency information into the background wavefield, can improve the convergence, as well as improve the accuracy and generalization ability a lot compared to conventional neural operator-based methods. 

We further compare their performance on the unseen velocities with 200 samples. As shown in Table~\ref{tab:pipeline_comp}, our method achieves the best accuracy compared to the conventional approaches with almost the same computational cost.
\begin{table}
    \centering
    \caption{The training cost and accuracy comparison between different approaches}
    \setlength{\tabcolsep}{0.035\columnwidth}
    \begin{tabular}{c|c|c|c}
    \toprule
        \multirow{2}{*}{Approaches} & \multicolumn{2}{c|}{Relative L2 error} & Training cost\\
        & Real & Imag. & (hours)\\
         \midrule
        Conventional in full out& 0.3290 & 0.3271& 15.714\\
        Conventional in residual out &0.6935 & 0.6943 & 15.771 \\
         Our method& \textbf{0.2598} & \textbf{0.2599} & 15.759\\
         \bottomrule
    \end{tabular}
    \label{tab:pipeline_comp}
\end{table}

\subsection{More realistic velocity models}
\begin{figure}
    \centering
    \includegraphics[width=\columnwidth]{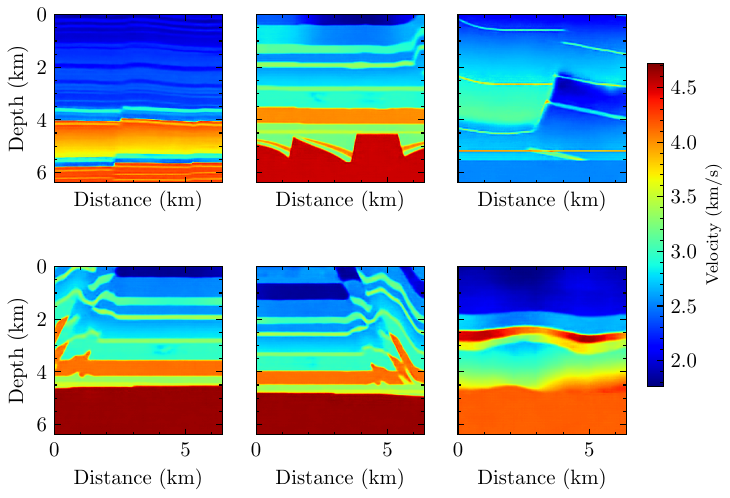}
    \caption{Samples of realistic velocity models used in the training.}
    \label{fig:real_velocities}
\end{figure}

\begin{figure}
    \centering
    \includegraphics[width=0.5\columnwidth]{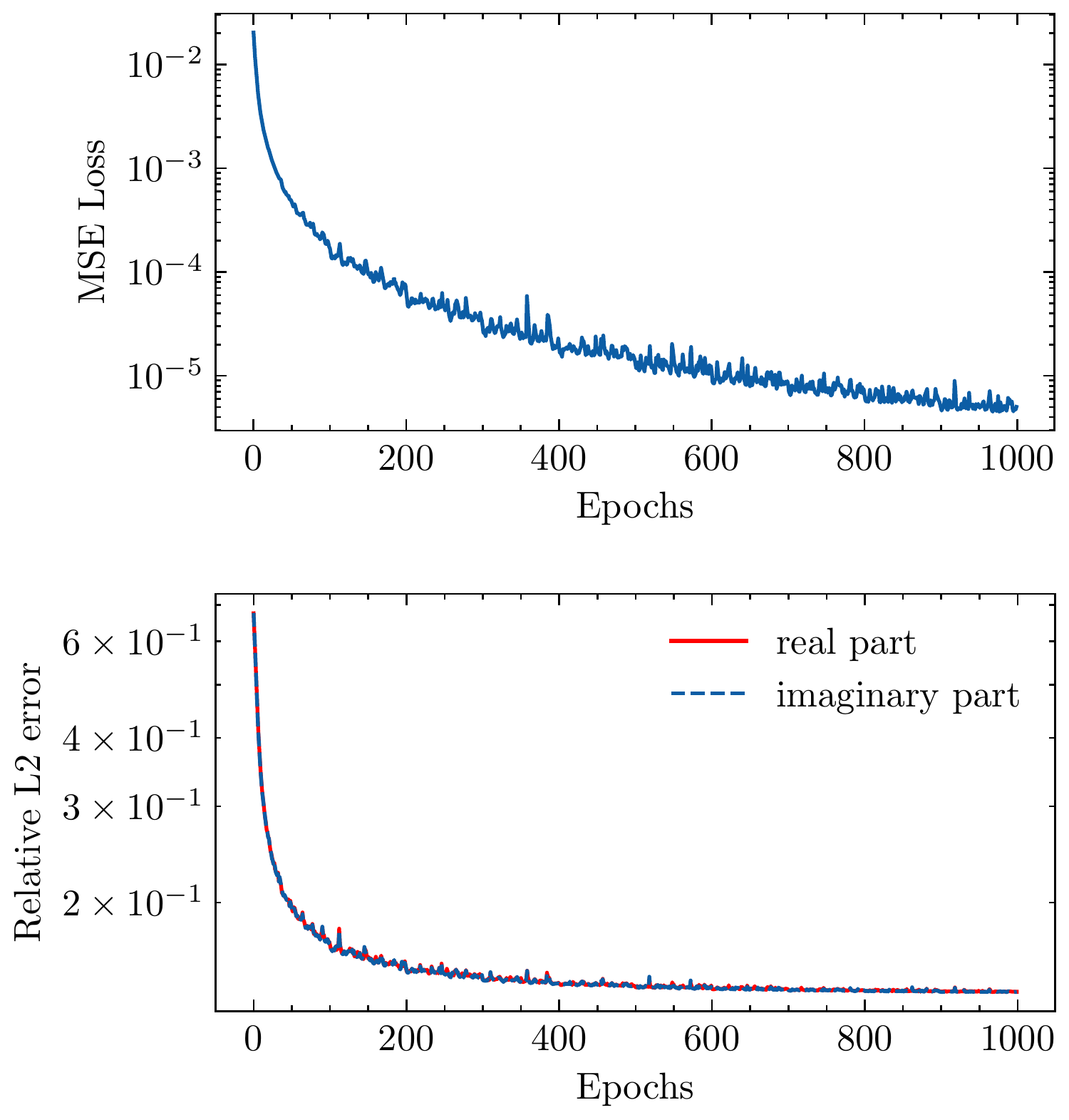}
    \caption{The training history for the realistic models test. \textbf{Top}: the training loss curve; \textbf{Bottom}: the validation error.}
    \label{fig:training_realistic}
\end{figure}

\begin{figure}
    \centering
    \includegraphics[width=1.0\textwidth]{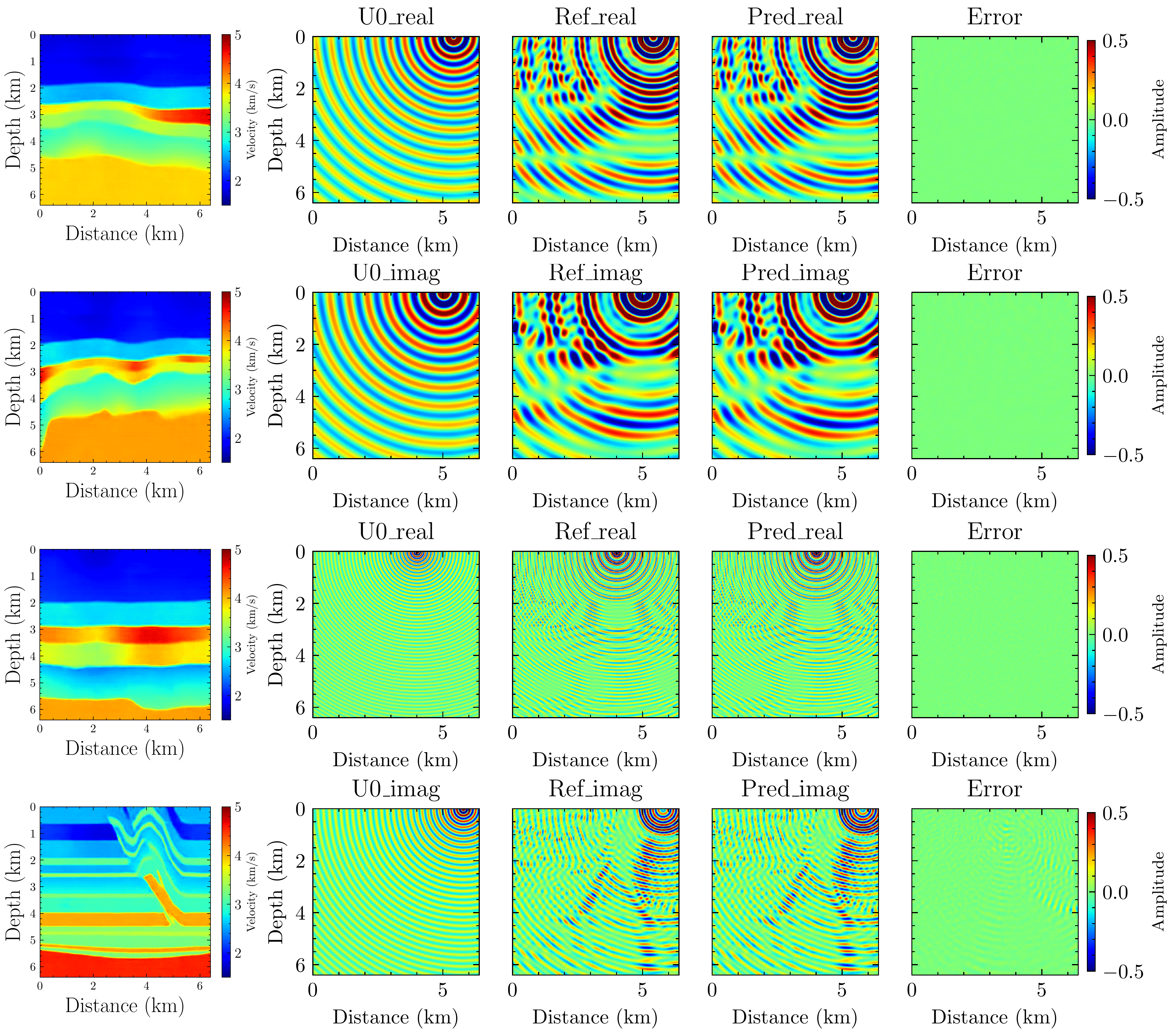}
    \caption{The results of the network predictions for four realistic velocity models in which the order of the plots are the same as in Figure \ref{fig:openfwi_results}. The frequencies for each forward modeling are 5.00 Hz, 3.93 Hz, 20.66 Hz, and 12.60 Hz from top to bottom.}
    \label{fig:real_results}
\end{figure}

\begin{figure}
    \centering
    \includegraphics[width=1.0\textwidth]{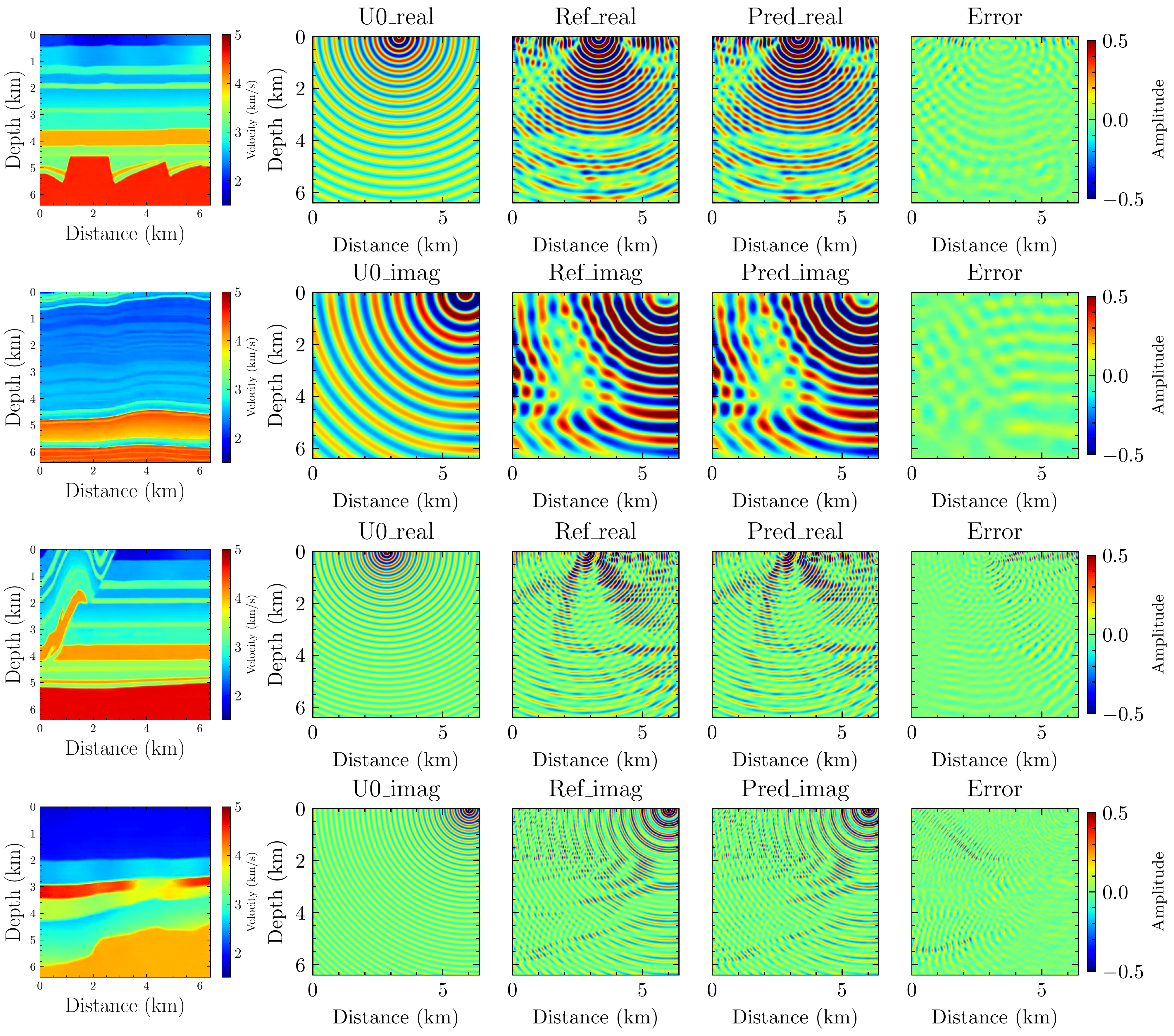}
    \caption{The results of the network predictions for four unseen realistic velocity models in which the order of the plots are the same as in Figure \ref{fig:openfwi_results}. The frequencies for each forward modeling are 7.19 Hz, 3.32 Hz, 11.98 Hz, and 13.76 Hz from top to bottom.}
    \label{fig:real_results_unseen}
\end{figure}

To further test the effectiveness of the proposed method, we choose more realistic velocity models to test, e.g., the combinations of Overthrust, Marmousi models, and so on.
We extracted 256×256 patches from the complete velocity models and applied shifting and reversal operations to generate an augmented set of velocity models. Figure~\ref{fig:real_velocities} shows samples of velocities in the training dataset.
The number of training samples is 5,400, while that for validation is 600. 
For each sample, the frequency was chosen between 3 Hz and 21 Hz, the velocity was randomly selected from the dataset, the source location was randomly positioned on the surface, and a constant background velocity of 2.8 km/s was used to generate the corresponding background and scattered wavefields.
Figure~\ref{fig:training_realistic} shows the training loss curve and the validation error at each epoch. 
We also train FNO up to 1000 epochs, and it converges well.
We similarly obtain the numerical reference results on unseen cases with a spatial interval of 0.025 km in both $x$ and $z$ directions and compare them with the predictions from the FNO.
Realistic models are considerably more complex than the ‘CurveVelA’ models, and the available velocity model samples do not fully capture the complete distribution of realistic models. Consequently, the generalization in realistic models is much more challenging. 
To demonstrate the effectiveness of our approach on complex models, we evaluate it under two scenarios. 
In the first scenario, we assess the generalization with respect to frequency and source location by using a velocity model from the training set while employing source locations and frequency configurations that were not encountered during training for the corresponding velocity model. 
The results, shown in Figure~\ref{fig:real_results}, indicate accurate predictions.
In the second scenario, we evaluate performance on unseen velocity models, with four representative examples presented in Figure~\ref{fig:real_results_unseen}. Although the predictions generally align with the reference numerical results thanks to the inclusion of similar velocity models in the training dataset, we observe that higher frequencies tend to introduce larger errors. This finding underscores that generalizing across different velocity models is more challenging than generalizing over variations in source and frequency configurations.
We contend that a larger set of training samples covering the entire distribution of velocity models is necessary to enhance the accuracy of neural operator-based simulations for complex velocity models.

\subsection{Generalization test with reference frequency}
\begin{figure}
    \centering
    \includegraphics[width=1.0\textwidth]{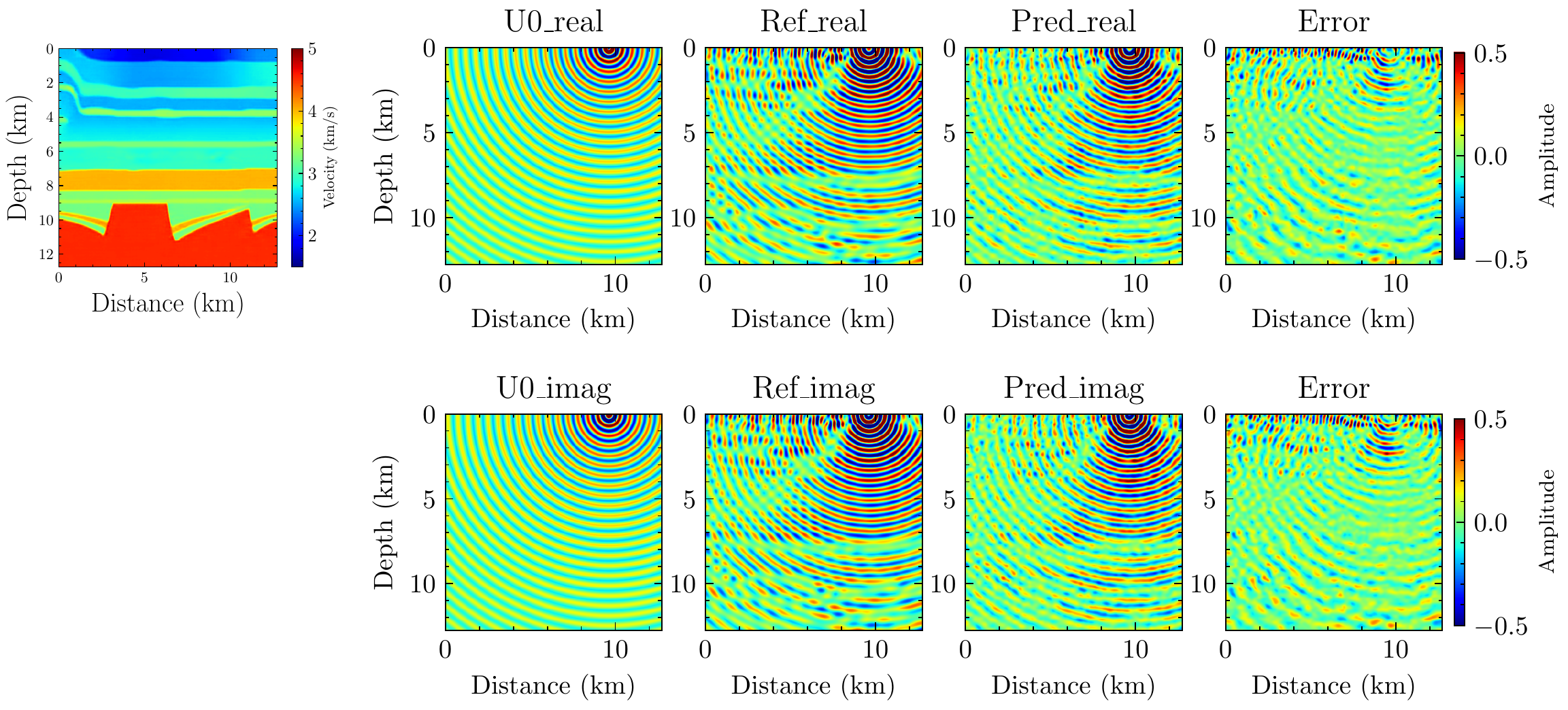}
    \caption{A generalization test on a realistic velocity model (Column 1) covering an area of 12.8$\times$12.8$km^2$. Column 2 shows the real and imaginary parts of the analytical background wavefield defined on 12.8$\times$12.8$km^2$ with an interval of 0.025 km in both $x$ and $z$ directions and with a frequency of 4.78 Hz; Column 3 represents the numerical solutions using optimal 9-point finite-difference method on a grid of 512 $\times$ 512 with a spatial interval of 0.025 $km$; Column 4 and Column 5 are the predictions using a conventional method and corresponding errors compared to the numerical solutions.}
    \label{fig:real_generalization}
\end{figure}

\begin{figure}
    \centering
    \includegraphics[width=0.6\textwidth]{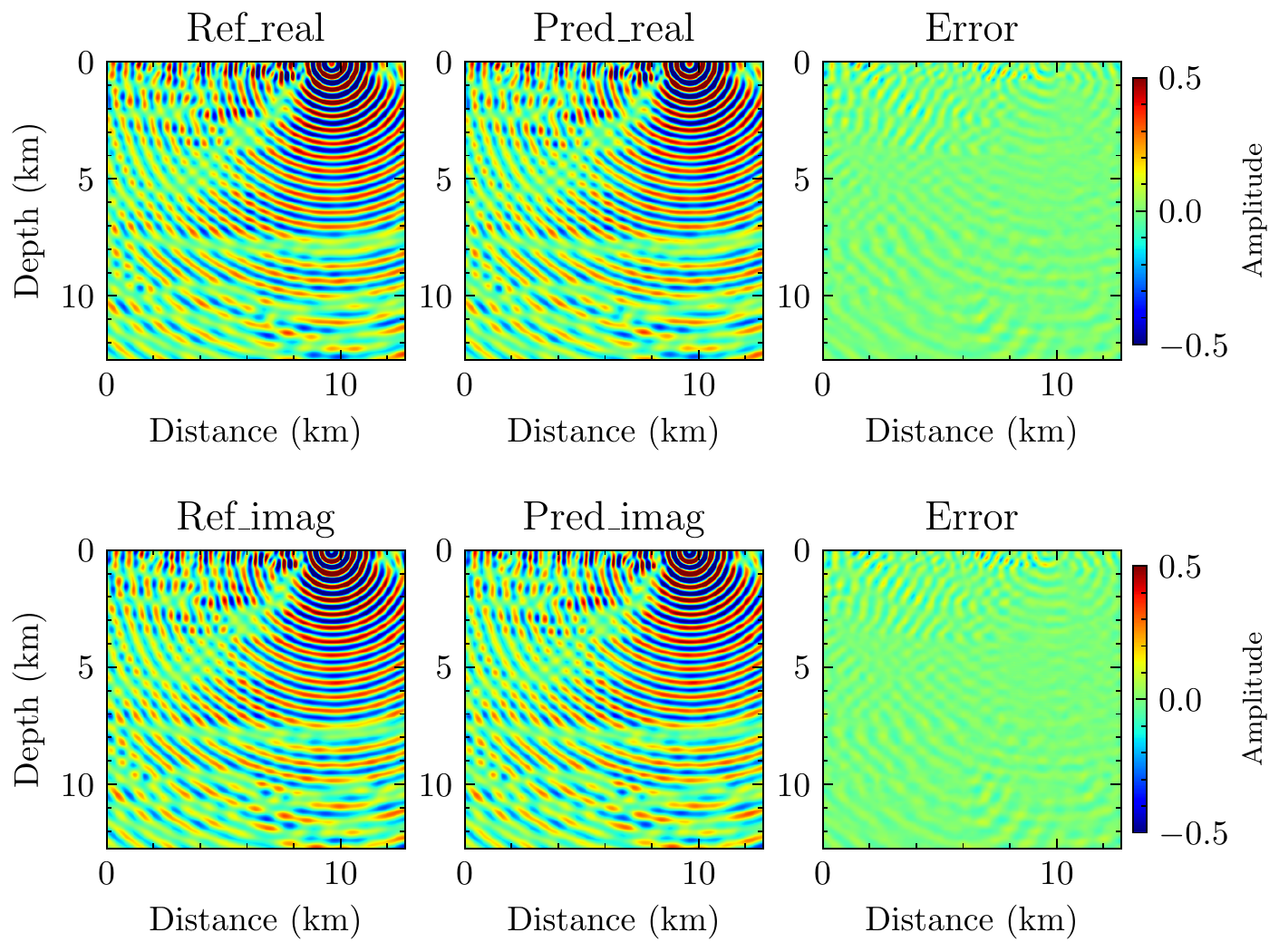}
    \caption{A generalization test on a realistic velocity model (the same velocity shown in Figure \ref{fig:real_generalization}) using our method. Columns 1, 2, and 3 show the numerical solutions, the predictions of our method, and the errors.}
    \label{fig:real_generalization_reference_way}
\end{figure}

While FNO supports zero-shot super-resolution simulation \citep{li_fourier_2020}, as shown in \cite{raonic_convolutional_2023}, it faces aliasing problems across different resolution scales, yielding large errors when we apply the pre-trained neural operator to predict wavefields for an unseen resolution.
Beyond super-resolution, where physical simulations are performed on the same spatial domain but with finer intervals, generalizing to larger domains is challenging. 
In this scenario, the spatial resolution remains consistent with the training data, but the overall coverage area expands.
So, we extend our testing to evaluate the proposed method's generalization capabilities on an unseen sample for a larger model, utilizing the concept of reference frequency. The neural network was initially trained on velocities covering an area of 6.4$\times$6.4 km$^2$ with a resolution of 256$\times$256.
We then conducted a generalization test on a velocity model spanning 12.8$\times$12.8 km$^2$, with the same spatial interval.
Due to the inherent resolution independence of the FNO, it seamlessly outputs the wavefield for the larger domain, which we refer to as 'direct prediction', as depicted in Figure~\ref{fig:real_generalization}.
Although the main components of the wavefield have been predicted by the FNO, their relative errors are still large. We calculate the relative L2 norm, which is the Euclidean distance from the prediction to the reference solutions divided by the Euclidean norm of the reference solutions, and the relative errors for direct predictions are 0.542 for the real part and 0.537 for the imaginary part of the wavefields.
It is hard to obtain a high-accuracy prediction for a large unseen domain directly. 
Crucially, leveraging the reference frequency concept \citep{huang_single_2022} allows us to upscale the 4.78 Hz wavefield in the large domain (12.8$\times$12.8 km$^2$) to an equivalent 9.56 Hz wavefield in the smaller domain (6.4$\times$6.4 km$^2$). 
This provides us with physical support to predict high-frequency wavefields for a large-scale velocity model rather than a direct prediction of high-resolution wavefields. 
As illustrated in Figure~\ref{fig:real_generalization_reference_way}, this scaled wavefield demonstrates enhanced accuracy compared to direct predictions. 
The relative errors calculated at the resolution of 256$\times$256 between the predictions of our method without interpolation and the reference results are 0.143 for the real part and 0.142 for the imaginary part of the wavefield. 
After bilinear interpolation of the prediction using our method, the errors at a resolution of 512$\times$512 are 0.171 for the real part and 0.170 for the imaginary part of the wavefield. Though the errors slightly increase, as shown in Figure \ref{fig:real_generalization_reference_way}, the predictions using our method are still close to the reference results.

\section{Discussions} 
\label{sec:discussion}
\begin{figure}
    \centering
    \includegraphics[width=0.4\textwidth]{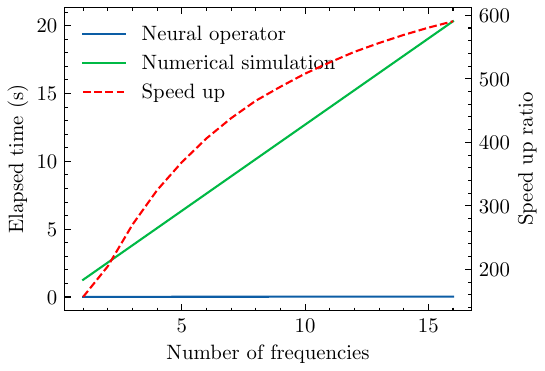}
    \caption{The efficiency comparison in terms of elapsed time for the learned wavefield solutions and the numerical finite-difference method on a resolution of 139$\times$139 with respect to different numbers of frequencies.}
    \label{fig:efficiency_plot}
\end{figure}

The proposed method has shown good performance on the OpenFWI and realistic velocity models. 
It frees us at inference time from numerical simulations, in which the computational complexity is high.
Taking the test on realistic models as an example, the numerical simulation for each velocity and frequency, executed in parallel on an Intel(R) Xeon(R) Gold 6230R CPU \atsymbol2.10GHz, at a resolution of 139$\times$139 requires 1.27 seconds, and that at a resolution of 256$\times$256 and 512$\times$512 require 2.79 and 8.64 seconds, respectively.
Using our method, the elapsed time on an NVIDIA A100 GPU (we do not fully utilize the GPU due to the relatively small input size and batch size) is 0.008 seconds for the resolution of 139$\times$139, 0.012 seconds for the resolution of 256$\times$256, and 0.031 seconds for the resolution of 512$\times$512. 
The main computationally intensive part of the proposed method is the training. 
In our experiments, it took 15.75 hours for training the neural operator for the "CurveVelA velocity models, and 29.25 hours for the realisitc velocity models.
Noteworthy, all of the training are performed offline. 
When trying to apply our method, indeed the neural operator-based wavefield solutions, on the downstream tasks such as reverse time migration or full waveform inversion, which requires massive simulations on different frequencies and source locations, the speed-up ratio of our method compared to conventional numerical methods will significantly increase, which will easily justify the expensive training.
Taking multiple frequencies simulations on a resolution of 139$\times$139 as an example, the elapsed times for learned wavefield solutions and numerical finite-difference method with respect to the number of frequencies are shown in Figure~\ref{fig:efficiency_plot}. 
Since the learned wavefield solutions make it easy to parallel the prediction without additional modification (like parallel computing required for the numerical method to handle multiple frequencies simultaneously), instead, we just increase the batch size of the input tensors.
It is obvious that within 16 simulations, the elapsed time for the learned wavefield solutions did not exceed 1 s, while the elapsed time for the numerical simulation without additional modification increased linearly. 
Due to the saturation of computational resources and memory bandwidth, the cost of neural operator-based wavefield solutions is slightly increased, causing a non-linear increase in the speed-up ratio.
In summary, at inference, it is much faster than the conventional numerical solvers.
Those speedups come from two main aspects: (1) we avoid repeated simulations; (2) we can easily utilize the latest hardware and software advances of accelerated computing \citep{azizzadenesheli_neural_2024} because we build the algorithm on machine learning.
However, we recognize the challenges associated with the preparation of data-label pairs required for training the neural operator, particularly as the resolution of training samples increases or when dealing with complex 3D cases, e.g., 3D viscoelastic anisotropic modeling. 
In addition, when dealing with high-resolution 3D simulation, the computational cost of FNO will remain a challenge. This might require light 3D neural networks such as 3D Unet \citep{wandel_teaching_2021} or LordNet3D \citep{huang_lordnet_2024}, which are efficient for 3D simulation.

\begin{figure}
    \centering
    \includegraphics[width=0.8\linewidth]{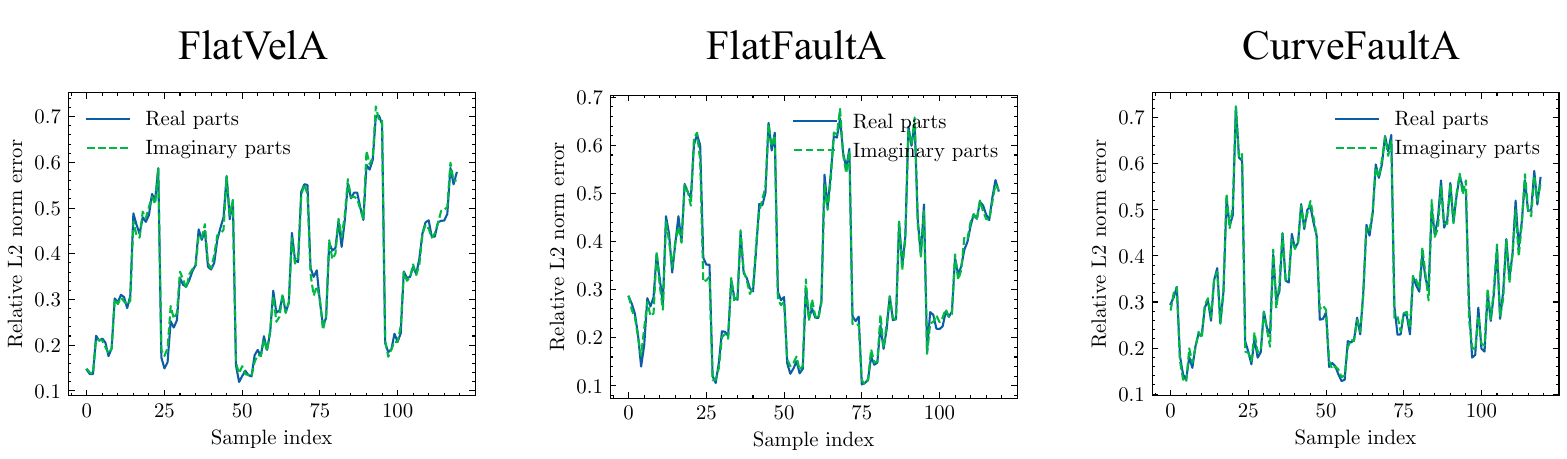}
    \caption{The relative L2 norm errors for each class of models. Regarding the sample index from 0 to 120, each of the 24 consecutive numbers represents a velocity model, and each of the three consecutive numbers represents a frequency. Then, each velocity model includes 8 frequencies, increasing in sequence, and each frequency value corresponds to three different shots.}
    \label{fig:ood_plot}
\end{figure}

\begin{figure}
    \centering
    \includegraphics[width=1.0\linewidth]{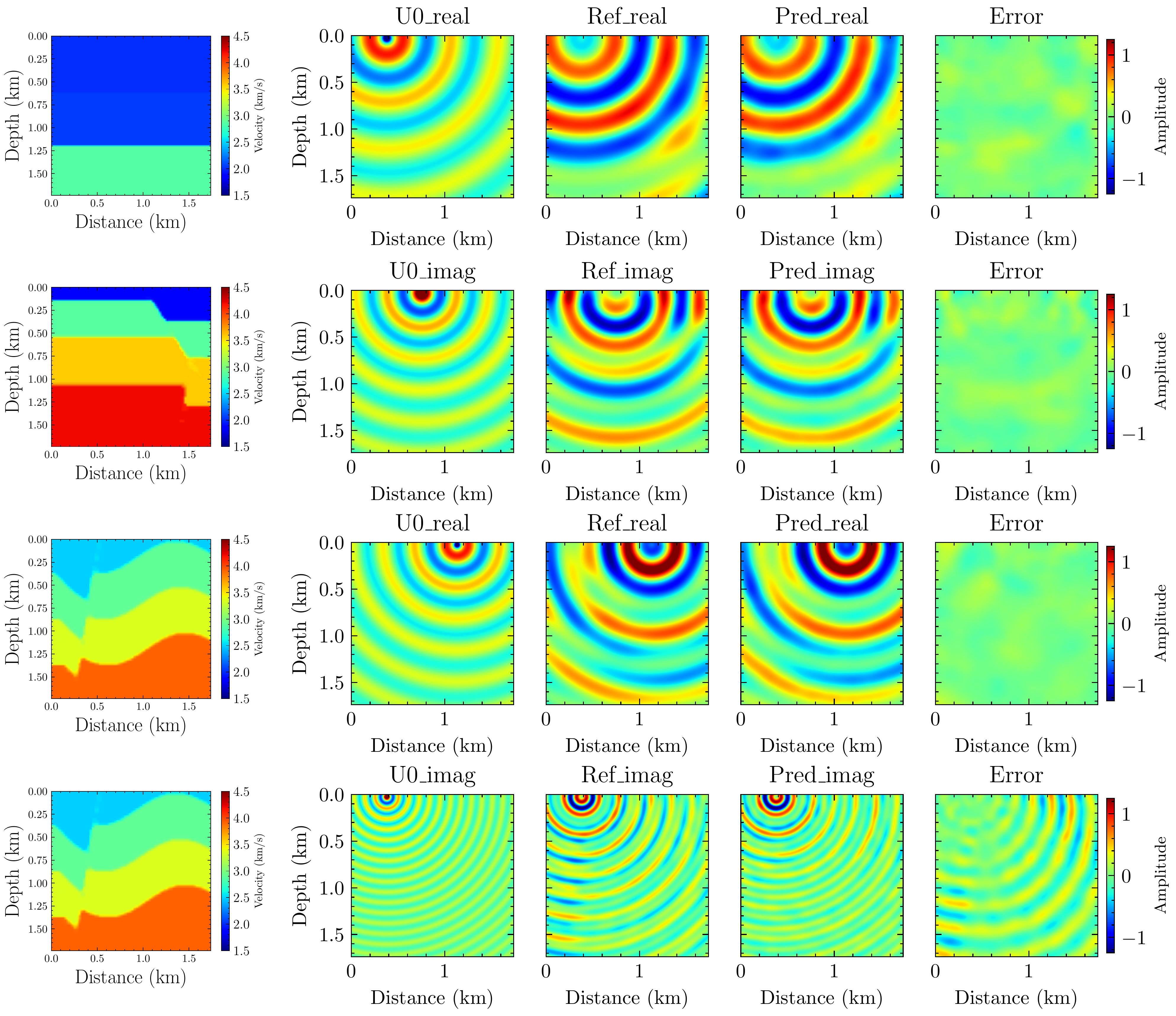}
    \caption{The samples of the out-of-distribution test. The first three rows are the samples on "FlatVelA", "FlatFaultA", and "CurveFaultA" for low-frequency simulation, while the fourth row is the worst case on "CurveFaultA" for high-frequency simulation.}
    \label{fig:ood_vis}
\end{figure}

In terms of generalization, taking the neural operator trained on "CurveVelA" as an example, we further evaluate its performance on other entirely different types of velocity models such as "FlatVelA" (no curvature in the velocity model), "FlatFaultA" (no curvature, but include faults), and "CurveFaultA" (with curvature and faults).
For each class, we test five velocity models; for each velocity model, we test eight frequencies; and for each frequency, we test three shots. 
We show the relative L2 norm error for each class in Figure~\ref{fig:ood_plot}.
The errors for high-frequency prediction are large, while those for low-frequency are low.
We show three good results and one poor case among these tests in Figure~\ref{fig:ood_vis}.
For those testing velocities with quite different features compared to the training datasets, our method can still generalize well for the low-frequency wavefield predictions. 
However, it fails to accurately predict high-frequency wavefields on the "CurveFaultA" velocity model.
The generalization capability to unseen velocities, hinges on the diversity and volume of the training samples.
While these challenges can be mitigated to some extent through physics-constrained learning, our focus in this paper is on introducing the neural operator with the background wavefield as input. 

\begin{figure}
    \centering
    \includegraphics[width=0.4\textwidth]{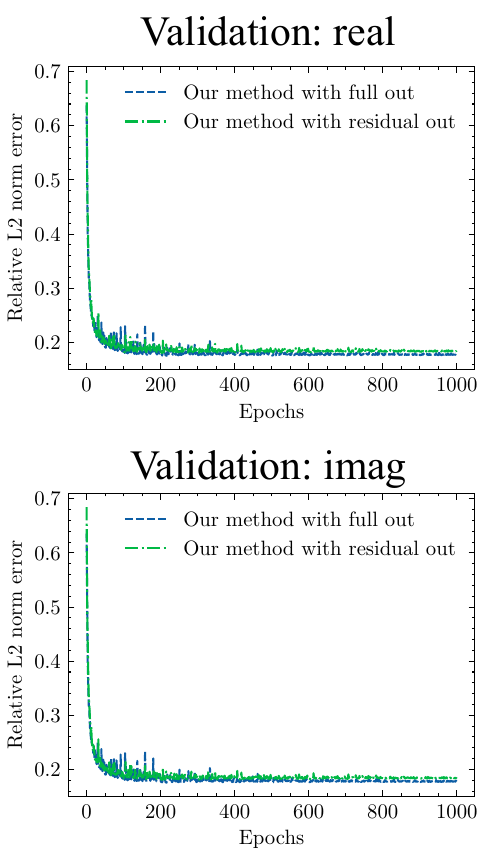}
    \caption{The comparison between FNO with scattered wavefield as output and that with full wavefield as output on the validation errors for the real and imaginary parts}
    \label{fig:fno_res_full_comp}
\end{figure}

Furthermore, the generalization prowess of our method across varied velocity models holds promising implications for applications in full waveform inversion (FWI), potentially eliminating the need for retraining and achieving real-time FWI. 
In our implementation, we decided to have the actual velocity as input and the scattered wavefield as output. However, the method can be equally applied by alternatively having the perturbation velocity (difference between the true and homogeneous background) as input and also the option of having the scattered or the full wavefield as output. 
Here, we take CurveVelA as an example to compare the difference between our method and our method with full wavefield output, shown in Figure~\ref{fig:fno_res_full_comp}.
Their performance is almost the same.
However, as mentioned earlier, the full wavefield can also be obtained by employing a residual connection between the input background wavefield and the scattered wavefield.

Finally, to further highlight the effectiveness of the proposed method on different neural network architectures, we compare the different approaches such as such as conventional
inputs with full wavefield outputs (marked as ”Conventional in
full out”), conventional inputs with scattered wavefield outputs
(marked as ”Conventional in residual out”), our method with full wavefield outputs, and our method with scattered wavefield outputs on different neural network architectures. 
We pick ResNet \citep{he_deep_2016} and Unet \citep{Wandel2020} for the study.
The results are shown in Tables~\ref{tab:pipeline_comp-resnet} and~\ref{tab:pipeline_comp-unet}.
The results demonstrate that our method is generally effective with other neural networks.
Regarding the type of output, we found that a residual output, given by the scattered wavefield here, will slightly improve the accuracy of the wavefield prediction when using U-Net while performing almost the same in ResNet and FNO. 
However, we found that for conventional input, which includes sub-optimal binary masks for the source location and constant channel for the frequency information, the residual output instead performs worse. 
We do not further investigate it in our paper and leave it for future investigation. 
Overall, these results indicate that the conventional, sub-optimal input is highly sensitive to the output type, resulting in inconsistent accuracy performance. 
In contrast, our proposed informative input, which includes the background wavefield and velocity as input, provides an additional benefit by making the neural operator-based simulation more robust to variations in the output type compared to the conventional approach.
\begin{table}
    \centering
    \caption{The training cost and accuracy comparison between different approaches, using ResNet as backbone}
    \setlength{\tabcolsep}{0.035\columnwidth}
    \begin{tabular}{c|c|c|c}
    \toprule
        \multirow{2}{*}{Approaches} & \multicolumn{2}{c|}{Relative L2 error} & Training cost\\
        & Real & Imag. & (hours)\\
         \midrule
        Conventional in full out& 0.7411& 0.7391& 11.956\\
        Conventional in residual out &0.9479 & 0.9517 & 11.912 \\
         Our method with full out& \textbf{0.5730} & 0.5671 & 11.968\\
         Our method with residual out& 0.5748& \textbf{0.5653} & 11.940\\
         \bottomrule
    \end{tabular}
    \label{tab:pipeline_comp-resnet}
\end{table}
\begin{table}
    \centering
    \caption{The training cost and accuracy comparison between different approaches, using Unet as backbone}
    \setlength{\tabcolsep}{0.035\columnwidth}
    \begin{tabular}{c|c|c|c}
    \toprule
        \multirow{2}{*}{Approaches} & \multicolumn{2}{c|}{Relative L2 error} & Training cost\\
        & Real & Imag. & (hours)\\
         \midrule
        Conventional in full out& 0.3050 & 0.3074& 2.394\\
        Conventional in residual out &0.4120 & 0.4160 & 2.417 \\
         Our method with full out& 0.1963 & 0.1957 & 2.363\\
         Our method with residual out& \textbf{0.1905} & \textbf{0.1892} & 2.375\\
         \bottomrule
    \end{tabular}
    \label{tab:pipeline_comp-unet}
\end{table}

\section{Conclusions}
We presented a novel implementation for predicting the wavefield using the Fourier Neural Operator (FNO) by incorporating the analytically derived background wavefield as input. As a result, the source location and frequency are embedded in the background wavefield, facilitating compact and effective feature engineering. This methodology enables efficient surrogate modeling across various velocities, frequencies, and source locations. Our tests on both the OpenFWI dataset and some realistic velocity models have demonstrated the high accuracy and effectiveness of the proposed method.
Additionally, incorporating a reference frequency strategy significantly enhances the accuracy of our model in generalizing to a larger, unseen domain, marking a substantial advancement in the field of frequency-domain wavefield modeling.

\begin{acknowledgments}
We thank KAUST and the DeepWave Consortium sponsors for their support and the SWAG group for the collaborative environment. This work utilized the resources of the Supercomputing Laboratory at KAUST, and we are grateful for that. 
\end{acknowledgments}

\bibliographystyle{gji}
\bibliography{opl}
  
\end{document}